
%
%
\magnification=\magstep2
\baselineskip=3ex plus 0.2ex minus 0.2ex
\hoffset=-0.6truecm
\voffset=-0.6truecm
\hsize=17.0truecm
\vsize=23.0truecm
\overfullrule=0pt
\def\dsas{$\Delta_{SAS}$}

%
%



\def\body                      
  {\beginparmode}              

\def\beginparmode{\endmode
  \begingroup \def\endmode{\par\endgroup}}
\let\endmode=\par
{\obeylines\gdef\
{}}
%
%

\def\figurecaptions
  {\leftline{\bf Figure Captions}
   \beginparmode
   \frenchspacing \parindent=0pt \leftskip=1.1cm
   \interlinepenalty=10000
   \parskip=8pt plus 3pt \everypar{\hangindent=\parindent}}

\gdef\figis#1{\indent\hbox to 0pt{\hss Fig.#1.~}} 
%
\def\refto#1{$^{#1)}$}          

\def\references       
  {\leftline{\bf References}
   \beginparmode
   \frenchspacing \parindent=0pt \leftskip=0.6cm
   \interlinepenalty=10000
   \parskip=8pt plus 3pt \everypar{\hangindent=\parindent}}

\gdef\refis#1{\indent\hbox to 0pt{\hss#1.~}} 

\gdef\journal#1, #2, #3, 1#4#5#6{      
        #1~ {\bf #2}, #3 (1#4#5#6)}   

\def\endreferences{\body}

\def\endfigurecaptions{\body}

\def\endpage                   
  {\vfill\eject}

\def\endpaper                  
  {\endmode\vfill\supereject}


\def\Ref#1{Ref.#1}                   
\def\Refs#1{Refs.#1}                 
\def\Nref#1{#1}                   

\def\ref#1{ref.#1}                   

\def\Figure#1{Figure\ #1}
\def\Figures#1{Figures #1}
\def\Fig#1{Fig.#1}
\def\Figs#1{Figs.#1}
\def\Nfig#1{#1}

\def\Equation#1{Equation\ (#1)}       
\def\Equations#1{Equations (#1)}       
\def\Eq#1{Eq.(#1)}                     
\def\Eqs#1{Eqs.(#1)}                   
\def\Neq#1{(#1)}       

\def\eq#1{eq.(#1)}                     
\def\eqs#1{eqs.(#1)}                   

\catcode`@=11
\newcount\tagnumber\tagnumber=0

\immediate\newwrite\eqnfile
\newif\if@qnfile\@qnfilefalse
\def\write@qn#1{}
\def\writenew@qn#1{}
\def\w@rnwrite#1{\write@qn{#1}\message{#1}}
\def\@rrwrite#1{\write@qn{#1}\errmessage{#1}}

\def\taghead#1{\gdef\t@ghead{#1}\global\tagnumber=0}
\def\t@ghead{}

\expandafter\def\csname @qnnum-3\endcsname
  {{\t@ghead\advance\tagnumber by -3\relax\number\tagnumber}}
\expandafter\def\csname @qnnum-2\endcsname
  {{\t@ghead\advance\tagnumber by -2\relax\number\tagnumber}}
\expandafter\def\csname @qnnum-1\endcsname
  {{\t@ghead\advance\tagnumber by -1\relax\number\tagnumber}}
\expandafter\def\csname @qnnum0\endcsname
  {\t@ghead\number\tagnumber}
\expandafter\def\csname @qnnum+1\endcsname
  {{\t@ghead\advance\tagnumber by 1\relax\number\tagnumber}}
\expandafter\def\csname @qnnum+2\endcsname
  {{\t@ghead\advance\tagnumber by 2\relax\number\tagnumber}}
\expandafter\def\csname @qnnum+3\endcsname
  {{\t@ghead\advance\tagnumber by 3\relax\number\tagnumber}}

\def\equationfile{%
  \@qnfiletrue\immediate\openout\eqnfile=\jobname.eqn%
  \def\write@qn##1{\if@qnfile\immediate\write\eqnfile{##1}\fi}
  \def\writenew@qn##1{\if@qnfile\immediate\write\eqnfile
    {\noexpand\tag{##1} = (\t@ghead\number\tagnumber)}\fi}
}

\def\callall#1{\xdef#1##1{#1{\noexpand\call{##1}}}}
\def\call#1{\each@rg\callr@nge{#1}}

\def\each@rg#1#2{{\let\thecsname=#1\expandafter\first@rg#2,\end,}}
\def\first@rg#1,{\thecsname{#1}\apply@rg}
\def\apply@rg#1,{\ifx\end#1\let\next=\relax%
\else,\thecsname{#1}\let\next=\apply@rg\fi\next}

\def\callr@nge#1{\calldor@nge#1-\end-}
\def\callr@ngeat#1\end-{#1}
\def\calldor@nge#1-#2-{\ifx\end#2\@qneatspace#1 %
  \else\calll@@p{#1}{#2}\callr@ngeat\fi}
\def\calll@@p#1#2{\ifnum#1>#2{\@rrwrite{Equation range #1-#2\space is
   bad.}
\errhelp{If you call a series of equations by the notation M-N, then M
  and N must be integers, and N must be greater than or equal to M.}}
 \else%
 {\count0=#1\count1=#2\advance\count1
 by1\relax\expandafter\@qncall\the\count0,%
  \loop\advance\count0 by1\relax%
    \ifnum\count0<\count1,\expandafter\@qncall\the\count0,%
  \repeat}\fi}

\def\@qneatspace#1#2 {\@qncall#1#2,}
\def\@qncall#1,{\ifunc@lled{#1}{\def\next{#1}\ifx\next\empty\else
  \w@rnwrite{Equation number \noexpand\(>>#1<<) has not been defined
             yet.}
  >>#1<<\fi}\else\csname @qnnum#1\endcsname\fi}

\let\eqnono=\eqno
\def\eqno(#1){\tag#1}
\def\tag#1$${\eqnono(\displayt@g#1 )$$}

\def\aligntag#1\endaligntag
  $${\gdef\tag##1\\{&(##1 )\cr}\eqalignno{#1\\}$$
  \gdef\tag##1$${\eqnono(\displayt@g##1 )$$}}

\def\eqalignno#1{\displ@y \tabskip\centering
  \halign to\displaywidth{\hfil$\displaystyle{##}$\tabskip\z@skip
    &$\displaystyle{{}##}$\hfil\tabskip\centering
    &\llap{$\displayt@gpar##$}\tabskip\z@skip\crcr
    #1\crcr}}

\def\displayt@gpar(#1){(\displayt@g#1 )}

\def\displayt@g#1 {\rm\ifunc@lled{#1}\global\advance\tagnumber by1
        {\def\next{#1}\ifx\next\empty\else\expandafter
        \xdef\csname @qnnum#1\endcsname{\t@ghead\number\tagnumber}\fi}%
  \writenew@qn{#1}\t@ghead\number\tagnumber\else
        {\edef\next{\t@ghead\number\tagnumber}%
        \expandafter\ifx\csname @qnnum#1\endcsname\next\else
        \w@rnwrite{Equation \noexpand\tag{#1} is a duplicate number.}
        \fi}%
  \csname @qnnum#1\endcsname\fi}

\def\ifunc@lled#1{\expandafter\ifx\csname @qnnum#1\endcsname\relax}

\let\@qnend=\end\gdef\end{\if@qnfile
\immediate\write16{Equation numbers written on []\jobname.EQN.}\fi
\@qnend}

\catcode`@=12
\callall\Equation
\callall\Equations
\callall\Eq
\callall\Eqs
\callall\Neq
\callall\eq
\callall\eqs

\catcode`@=11
\newcount\r@fcount \r@fcount=0
\newcount\r@fcurr
\immediate\newwrite\reffile
\newif\ifr@ffile\r@ffilefalse
\def\w@rnwrite#1{\ifr@ffile\immediate\write\reffile{#1}\fi\message{#1}}

\def\writer@f#1>>{}
\def\referencefile{
  \r@ffiletrue\immediate\openout\reffile=\jobname.ref%
  \def\writer@f##1>>{\ifr@ffile\immediate\write\reffile%
    {\noexpand\refis{##1} = \csname r@fnum##1\endcsname = %
     \expandafter\expandafter\expandafter\strip@t\expandafter%
     \meaning\csname r@ftext\csname r@fnum##1\endcsname\endcsname}\fi}%
  \def\strip@t##1>>{}}

\def\citeall#1{\xdef#1##1{#1{\noexpand\cite{##1}}}}

\def\cite#1{\each@rg\citer@nge{#1}}     

\def\each@rg#1#2{{\let\thecsname=#1\expandafter\first@rg#2,\end,}}
\def\first@rg#1,{\thecsname{#1}\apply@rg}        
\def\apply@rg#1,{\ifx\end#1\let\next=\relax      
\else,\thecsname{#1}\let\next=\apply@rg\fi\next} 

\def\citer@nge#1{\citedor@nge#1-\end-}           
\def\citer@ngeat#1\end-{#1}
\def\citedor@nge#1-#2-{\ifx\end#2\r@featspace#1  
  \else\citel@@p{#1}{#2}\citer@ngeat\fi}        
\def\citel@@p#1#2{\ifnum#1>#2{\errmessage{Reference range #1-#2\space
  is bad.}
    \errhelp{If you cite a series of references by the notation M-N,
             then M and N must be integers, and N must be greater than
             or equal to M.}}\else%
 {\count0=#1\count1=#2\advance\count1
 by1\relax\expandafter\r@fcite\the\count0,%
  \loop\advance\count0 by1\relax
    \ifnum\count0<\count1,\expandafter\r@fcite\the\count0,%
  \repeat}\fi}

\def\r@featspace#1#2 {\r@fcite#1#2,}    
\def\r@fcite#1,{\ifuncit@d{#1}          
    \expandafter\gdef\csname r@ftext\number\r@fcount\endcsname%
    {\message{Reference #1 to be supplied.}\writer@f#1>>#1 to be
 supplied.\par
     }\fi%
  \csname r@fnum#1\endcsname}

\def\ifuncit@d#1{\expandafter\ifx\csname r@fnum#1\endcsname\relax%
\global\advance\r@fcount by1%
\expandafter\xdef\csname r@fnum#1\endcsname{\number\r@fcount}}

\def\regi#1{\each@rg\regir@nge{#1}}     

\def\regir@nge#1{\regidor@nge#1-\end-}           
\def\regidor@nge#1-#2-{\ifx\end#2\rr@featspace#1  
  \else\citel@@p{#1}{#2}\citer@ngeat\fi}        
\def\citel@@p#1#2{\ifnum#1>#2{\errmessage{Reference range #1-#2\space
  is bad.}
    \errhelp{If you cite a series of references by the notation M-N,
             then M and N must be integers, and N must be greater than
             or equal to M.}}\else%
 {\count0=#1\count1=#2\advance\count1
 by1\relax\expandafter\r@fcite\the\count0,%
  \loop\advance\count0 by1\relax
    \ifnum\count0<\count1,\expandafter\r@fcite\the\count0,%
  \repeat}\fi}

\def\rr@featspace#1#2 {\rr@fcite#1#2,}    
\def\rr@fcite#1,{\ifuncit@d{#1}          
    \expandafter\gdef\csname r@ftext\number\r@fcount\endcsname%
    {\message{Reference #1 to be supplied.}\writer@f#1>>#1 to be
 supplied.\par
     }\fi%
     }

\let\r@fis=\refis                       
\def\refis#1#2#3\par{\ifuncit@d{#1}     
    \w@rnwrite{Reference #1=\number\r@fcount\space is not cited up to
 now.}\fi%
  \expandafter\gdef\csname r@ftext\csname r@fnum#1\endcsname\endcsname%
  {\writer@f#1>>#2#3\par}}

\def\r@ferr{\endreferences\errmessage{I was expecting to see
\noexpand\endreferences before now;  I have inserted it here.}}
\let\r@ferences=\references
\def\references{\r@ferences\def\endmode{\r@ferr\par\endgroup}}

\let\endr@ferences=\endreferences
\def\endreferences{\r@fcurr=0
  {\loop\ifnum\r@fcurr<\r@fcount
    \advance\r@fcurr by
 1\relax\expandafter\r@fis\expandafter{\number\r@fcurr}%
    \csname r@ftext\number\r@fcurr\endcsname%
  \repeat}\gdef\r@ferr{}\endr@ferences}


\let\r@fend=\endpaper\gdef\endpaper{\ifr@ffile
\immediate\write16{Cross References written on []\jobname.REF.}
\fi\r@fend}

\catcode`@=12

\citeall\refto          
\citeall\Ref            %
\citeall\Refs           %
\citeall\Nref           %

\catcode`@=11
\newcount\f@gcount \f@gcount=0
\newcount\f@gcurr
\immediate\newwrite\figfile
\newif\iff@gfile\f@gfilefalse
\def\w@fnwrite#1{\iff@gfile\immediate\write\figfile{#1}\fi\message{#1}}

\def\writef@g#1>>{}
\def\figurefile{
  \f@gfiletrue\immediate\openout\figfile=\jobname.fig%
  \def\writef@g##1>>{\iff@gfile\immediate\write\figfile%
    {\noexpand\figis{##1} = \csname f@gnum##1\endcsname = %
     \expandafter\expandafter\expandafter\fstrip@t\expandafter%
     \meaning\csname f@gtext\csname f@gnum##1\endcsname\endcsname}\fi}%
  \def\fstrip@t##1>>{}}

\def\fciteall#1{\xdef#1##1{#1{\noexpand\fcite{##1}}}}
\def\fcite#1{\each@fg\fciter@nge{#1}}     

\def\each@fg#1#2{{\let\thecsname=#1\expandafter\first@fg#2,\end,}}
\def\first@fg#1,{\thecsname{#1}\apply@fg}        
\def\apply@fg#1,{\ifx\end#1\let\next=\relax      
\else,\thecsname{#1}\let\next=\apply@fg\fi\next} 

\def\fciter@nge#1{\fcitedor@nge#1-\end-}           
\def\fciter@ngeat#1\end-{#1}
\def\fcitedor@nge#1-#2-{\ifx\end#2\f@geatspace#1  
  \else\fcitel@@p{#1}{#2}\fciter@ngeat\fi}        
\def\fcitel@@p#1#2{\ifnum#1>#2{\errmessage{Figure range #1-#2\space
  is bad.}
    \errhelp{If you cite a series of figures by the notation M-N,
             then M and N must be integers, and N must be greater than
             or equal to M.}}\else%
 {\count0=#1\count1=#2\advance\count1
 by1\relax\expandafter\f@gcite\the\count0,%
  \loop\advance\count0 by1\relax
    \ifnum\count0<\count1,\expandafter\f@gcite\the\count0,%
  \repeat}\fi}

\def\f@geatspace#1#2 {\f@gcite#1#2,}    
\def\f@gcite#1,{\fifuncit@d{#1}          
    \expandafter\gdef\csname f@gtext\number\f@gcount\endcsname%
    {\message{Figure #1 to be supplied.}\writef@g#1>>#1 to be
 supplied.\par
     }\fi%
  \csname f@gnum#1\endcsname}

\def\fifuncit@d#1{\expandafter\ifx\csname f@gnum#1\endcsname\relax%
\global\advance\f@gcount by1%
\expandafter\xdef\csname f@gnum#1\endcsname{\number\f@gcount}}

\let\f@gis=\figis                       
\def\figis#1#2#3\par{\fifuncit@d{#1}     
    \w@fnwrite{Figure #1=\number\f@gcount\space is not cited up to
 now.}\fi%
  \expandafter\gdef\csname f@gtext\csname f@gnum#1\endcsname\endcsname%
  {\writef@g#1>>#2#3\par}}

\def\f@gerr{\endfigurecaptions\errmessage{I was expecting to see
\noexpand\endfigurecaptions before now;  I have inserted it here.}}
\let\f@gerences=\figurecaptions
\def\figurecaptions{\f@gerences\def\fendmode{\f@gerr\par\endgroup}}

\let\endf@gerences=\endfigurecaptions
\def\endfigurecaptions{\f@gcurr=0
  {\loop\ifnum\f@gcurr<\f@gcount
    \advance\f@gcurr by
 1\relax\expandafter\f@gis\expandafter{\number\f@gcurr}%
    \csname f@gtext\number\f@gcurr\endcsname%
  \repeat}\gdef\f@gerr{}\endf@gerences}


\let\f@gend=\fendpaper\gdef\fendpaper{\iff@gfile
\immediate\write16{Cross Figures written on []\jobname.REF.}
\fi\f@gend}

\catcode`@=12

\fciteall\Fig            
\fciteall\Figs           %
\fciteall\Figure         %
\fciteall\Figures        %
\fciteall\Nfig           %


\medskip
\centerline
{\bf Excitation Spectrum of a Double Quantum Well System}
\par
\centerline
{\bf in a Strong Magnetic Field}
\bigskip
\centerline
{Osamu NARIKIYO and Daijiro YOSHIOKA}
\medskip
\centerline
{\it Institute of Physics, University of Tokyo}
\par
\centerline
{\it Komaba, Meguro-ku, Tokyo 153, Japan}
\bigskip
The collective charge-excitation spectrum of
a double quantum well system in a strong magnetic field
is obtained within the random phase approximation.
Correction to the spectrum coming from the finiteness of the
magnetic field is calculated up to the first order in inverse of
the magnetic field.
Dependencies on the magnetic field of (i) the velocity of long wave
length excitation and (ii) the phase boundary of the quantum Hall state
signaled by the softening of the roton minimum of the spectrum
are discussed.

Keywords: quantum Hall effect, two dimensional electrons, strong magnetic
field, double quantum well, excitonic state

\regi{BJPW} \regi{SJSSG} \regi{SJSEHS}
\regi{SESST} \regi{EBPWH} \regi{SJSES}
\def\refEX{$^{\Nref{BJPW}-}$\refto{SJSES}}
\regi{RPM} \regi{RHV}
\regi{F} \regi{MPB} \regi{B} \regi{HXDSZ} \regi{CQl} \regi{CQp}
\def\refTH{$^{\Nref{RPM}-}$\refto{CQp}}
\bigskip
\noindent
{\bf 1. Introduction}
\taghead{1.}
\medskip
  Recently the experiments on the double quantum well system
in a strong magnetic field\refEX
have attracted great interests.
  Several theoretical calcula-\break tions\refTH
on this system have also been performed which include an attempt for
a field theoretical interpretation.\refto{WZ}
The present system is characterized by three parameters: $\nu$, $d$ and
\dsas.
We define the filling factor $\nu$ as a total filling factor,
namely sum of the Landau level filling factors of the two wells.
The separation between the two wells is denoted by $d$.
In the presence of the tunnelling between the two wells, symmetric and
antisymmetric combination of the single electron states in each well
become the eigenstates.
These eigenstates have energies different by an amount
\dsas.
Needless to say that $d$ cannot be varied freely in the real system,
and that \dsas ~is governed by $d$.
However, as a theoretical model it is customary
that we treat these as
free parameters to gain full understanding of the system.
It has been clarified that depending on these parameters
various  ground states
and the excitation spectrum can be realized.

In this paper we consider only systems with $\nu=1$.
Let us summarize previous investigations in this case.
Most of the investigations have been done in the quantum limit,
namely only the lowest Landau level is taken into account.
At $d=0$ and \dsas$=0$ the ground sate is almost equivalent
to the single well $\nu=1$ state, where each single electron
states in the lowest Landau level are occupied.
However, there is an important difference:
In the present case there are internal degree of freedom
or SU(2) symmetry\refto{RHV} for each Landau orbits,
namely each electron in a given Landau orbit has a choice to
be in any of the two wells or any linear combination of them,
which of course include symmetric and antisymmetric combinations.
Thus there is a huge degeneracy in this case.
These states can also be considered as  excitonic states.\refto{F,YM}
Namely we interprete an unoccupied Landau orbit in one of the wells as a hole.
Then this hole is always bound with an electron
in the same orbit of the other well to form an exciton.
The ground state of these free hard-core bosons is the excitonic state.
In this case the Goldstone mode makes the excitation spectrum gapless
and quadratic\refto{RPM,RHV} at small momentum.

When \dsas ~becomes finite this local SU(2) symmetry is broken:
only the symmetric states are favored.
The ground state at $d=0$ is still equivalent to the $\nu=1$ state,
and can be considered as an excitonic state, but the spectrum now
has a gap of the order of \dsas.
The ground state is unique, and the integer quantum Hall effect is
expected to be observed in this case.\refto{MPB}
We call this state as fully occupied symmetric (FOS) state hereafter.

The other parameter, $d$, has different effects.
When $d$ becomes finite, the inter-well Coulomb interaction between electrons
becomes weaker than the intra-well interaction.
Due to this difference the symmetric and antisymmetric states
are mixed up.\refto{MPB}
While the electron densities in each well are not fixed at $d=0$,
finite $d$ prefers the filling factors in each well to be fixed
at $1/2$.
Thus unique ground state is chosen among the degenerate excitonic states.
As long as \dsas=0 the excitation spectrum for this excitonic state
remains gapless even for finite $d$.
The long wave length excitation has a linear dispersion\refto{RHV,F}
which is nothing but density waves in each well
mutually out of phase.\refto{F}
Strictly speaking, the ground state is not the FOS state, when \dsas=0.
However, FOS is a good approximation when $d$ is small,
since the average filling factor
in each well is 1/2.

When $d$ becomes large enough, inter-well correlation
becomes less important.
In the excitonic state the inter-well and intra-well correlation have the same
strength.
Thus at large $d$  the excitonic state ceases to be the ground state.
The numerical investigations at \dsas=0 ~have shown that the excitonic state
is a good approximation to the ground state only for $d/l<1$, where
$l=\sqrt{c/eB}$ is the magnetic length.\refto{YM,YMG}
Corresponding to this
as $d$ becomes large the spectrum develops a minimum around
a wavenumber of the order of $1/l$.
This softening signals the instability of the excitonic state:
At around $d=l$ where the excitonic state ceases to be the ground
state, the minimum touches the zero energy axis.

MacDonald et al\refto{MPB} used this complete softening of
the excitation spectrum
to locate the phase boundary of the FOS state
for a system where both $d$ and \dsas are finite.
Since the existence of the energy gap in FOS state is guaranteed by the finite
\dsas,
they identified FOS as a state where quantum Hall effect is observed.
The nature of the ground state in region beyond the boundary
realized at larger $d$ has not been clarified.
MacDonald et al assumed that the ground state was gapless there
and would not show the quantum Hall effect.
This boundary qualitatively agrees with what experiments have shown.
There is a suggestion that the larger $d$ side of the boundary
is a kind of CDW state.\refto{B,CQl,CQp}
However, since the investigation has been done by Hartree-Fock
approximation, the true nature of the ground state would still be
an open question.

  The above mentioned calculations, however,
correspond to the strong magnetic field limit
taking only the contribution of the lowest Landau level into account and
do not always correspond to the experimental situations.
  In this paper
we take the contributions of higher Landau levels into account
within the random phase approximation
and obtain the magnetic field correction of the excitation spectrum.
We discuss the phase boundary of the excitonic ground states.
  We take a unit of $ \hbar = 1$.
\bigskip
\noindent
{\bf 2. Formulation}
\taghead{2.}
\medskip
  We study a double quantum well system in a strong magnetic field
using the model where the magnetic field is applied in the $z$ direction
and electrons are located in two planes perpendicular to the $z$ axis
which are separated by $d$.
  We first derive the effective hamiltonian
where contributions of higher Landau levels
are projected onto the lowest Landau level
and calculate the collective charge-excitation spectrum
within the random phase approximation.
\par
  In order to describe an electronic state in a magnetic field
we take an representation in the Landau gauge, ${\vec A}=(0,Bx,0)$;
the electron annihilation operator, $\Psi({\vec r},z)$
in the coordinate space is expressed as
$$
\Psi({\vec r},z)
= \sum_{k n \alpha}
\psi_k^{(n)}({\vec r}) \chi_\alpha (z) c_{k \alpha}^{(n)},
\eqno(wf0)
$$
where
${\vec r}$ denotes the vector in the $x$-$y$ plane.
In this equation $\psi_k^{(n)}({\vec r})$ is the two-dimensional wave function
given by
$$
\psi_k^{(n)}({\vec r})
= {1 \over {\sqrt L}} {\rm e}^{{\rm i}ky} \phi_k^{(n)}(x),
\eqno(wf1)
$$
$$
\phi_k^{(n)}(x)
= {1 \over \sqrt{{\sqrt \pi} l 2^n n!}}
{\rm e}^{-(x+kl^2)^2/2l^2} H_n[(x+kl^2)/l],
\eqno(wf2)
$$
with $H_n(x)$ being the $n$-th Hermite polynomial.
$\chi_\alpha (z)$ is the wave function in the $z$ direction
where $\alpha=s$  or $a$ denotes the symmetric ($s$) or the antisymmetric ($a$)
state.
The operator
$c_{k \alpha}^{(n)}$ annihilates an electron in the state
in the $n$-th Landau level with one-dimensional momentum $k$
and symmetry $\alpha$.
  With this basis the effective hamiltonian
first introduced by Paquet, Rice and Ueda\refto{PRU} is given in the
following form:
$$
\eqalignno{
H & = H_t + H_{int} \cr
    & H_t = {\Delta_{SAS} \over 2} \sum_k
           (c_{ka}^{(0)^\dagger} c_{ka}^{(0)}
          - c_{ks}^{(0)^\dagger} c_{ks}^{(0)}) & (Htun) \cr
    & H_{int} = \sum_{q_y k_1 k_2} \sum_{\alpha_1 \alpha_2 \alpha_3 \alpha_4}
                 V_{\alpha_1 \alpha_2 \alpha_3 \alpha_4}(q_y;k_1-k_2) \cr
    & \hskip4.50truecm \times
                 c_{k_1-q_y/2,\alpha_1}^{(0)^\dagger}
                 c_{k_2+q_y/2,\alpha_2}^{(0)^\dagger}
                 c_{k_2-q_y/2,\alpha_3}^{(0)}
                 c_{k_1+q_y/2,\alpha_4}^{(0)}, & (Hint)
           }
$$
where processes in the higher Landau levels are projected
onto the lowest Landau level within the second order perturbation
in terms of the inverse of the magnetic field.
The matrix element $V_{\alpha_1 \alpha_2 \alpha_3 \alpha_4}(q_y;k)$
include the effect of higher levels.
This $H_{int}$ can be rewritten conveniently, if we introduce
a charge density operator in the Landau gauge by
$$
\rho_{\alpha_1 \alpha_2}({\vec k}) = \sum_p
{\rm e}^{-{\rm i}k_x p l^2}
c_{p+k_y/2,\alpha_1}^{(0)^\dagger} c_{p-k_y/2,\alpha_2}^{(0)}.
\eqno(rho)
$$
Then
$$
H_{int}
= {1 \over 2L^2} \sum_{\vec q}
  \sum_{\alpha_1 \alpha_2 \alpha_3 \alpha_4}
  v_{\alpha_1 \alpha_2 \alpha_3 \alpha_4}^{eff}({\vec q})
  \rho_{\alpha_1 \alpha_2}({\vec q}) \rho_{\alpha_3 \alpha_4}(-{\vec q}),
\eqno(Hrhorho)
$$
where
$$
  v_{\alpha_1 \alpha_2 \alpha_3 \alpha_4}^{eff}({\vec q})
= f_{\alpha_1 \alpha_2 \alpha_3 \alpha_4}^{(0,0)}({\vec q})
+ g_{\alpha_1 \alpha_2 \alpha_3 \alpha_4}({\vec q})
\eqno(veff)
$$
and
@@$$V_{\alpha_1 \alpha_2 \alpha_3 \alpha_4}(q_y;k_1-k_2)
      ={1 \over 2L^2} \sum_{q_x}
      v^{eff}_{\alpha_1 \alpha_2 \alpha_3 \alpha_4}({\vec q})
      {\rm e}^{{\rm i}q_x(k_1-k_2)l^2}  .
      \eqno(relation)
    $$
In \Eq{veff} the first and the second terms give the contribution from
the lowest Landau level and the higher Landau levels, respectively,
and they are given as
$$
f_{\alpha_1 \alpha_2 \alpha_3 \alpha_4}^{(n,m)}({\vec q})
= v_{\alpha_1 \alpha_2 \alpha_3 \alpha_4}(q)
  A^{(n,m)}({\vec q}) {A^{(n,m)}}^\ast ({\vec q}),
\eqno(f)
$$
and
$$
\eqalignno{
g_{\alpha_1 \alpha_2 \alpha_3 \alpha_4}({\vec Q})
= - {1 \over 2L^2} \sum_{\vec P} & \sum_{\beta_1 \beta_2} \sum_n
    {\rm e}^{-{\rm i}(P_x Q_y - P_y Q_x) l^2}
    {1 \over 2 n \omega_c} \cr
\times
  [ & f_{\alpha_1 \alpha_2 \beta_1 \beta_2}^{(0,n)}
       ({{\vec P}+{\vec Q} \over 2})
\cdot f_{\beta_1 \beta_2 \alpha_3 \alpha_4}^{(n,0)}
       ({{\vec P}-{\vec Q} \over 2}) \cr
  + & f_{\alpha_1 \alpha_2 \beta_2 \beta_1}^{(0,n)}
       ({{\vec P}+{\vec Q} \over 2})
\cdot f_{\beta_1 \beta_2 \alpha_3 \alpha_4}^{(n,0)}
       ({{\vec Q}-{\vec P} \over 2}) ],
&(g)
           }
$$
where $\omega_c$ is the cyclotron frequency given by
$\omega_c=eB/m^\ast c$ with $m^\ast$ being the electron mass.
Here $v_{\alpha_1 \alpha_2 \alpha_3 \alpha_4}(q)$ is the
Coulomb matrix elements integrated in the $z$-direction:
$$
\eqalignno{
v_{\alpha_1 \alpha_2 \alpha_3 \alpha_4}(q)
  = {2\pi e^2 \over \epsilon q}
  \int_{-\infty}^\infty {\rm d}z_1& \int_{-\infty}^\infty  {\rm d}z_2
   {\rm e}^{-q|z_1-z_2|} \cr
  &  \times
     \chi_{\alpha_1} (z_1) \chi_{\alpha_2} (z_2)
     \chi_{\alpha_3} (z_2) \chi_{\alpha_4} (z_1) , & (v)
           }
$$
where $\epsilon$ is the dielectric constant.
In the actual calculation we choose the simplest form for $\chi_{\alpha}$:
$$
\chi_{s(a)}(z) = {{1}\over{\sqrt{2}}} [\phi_{+}(z) \pm \phi_{-}(z)],
\eqno(chi)
$$
where
$$
\phi^2_{\pm}(z)= \delta(z \pm {d\over 2}) .
\eqno(delta)
$$
Then
$$
v_{sasa}(q) = {\pi e^2 \over \epsilon q}
              \{ 1 - {\rm e}^{-qd} \}
\eqno(vsasa)
$$
and
$$
v_{ssss}(q) =
v_{ssaa}(q) = {\pi e^2 \over \epsilon q}
              \{ 1 + {\rm e}^{-qd} \} .
\eqno(vssss)
$$
The matrix element of the density operator $A^{(m,n)}(\vec q)$ in
\eq{f} is given as\refto{KMH}
$$
\eqalign{
A^{(m,n)}({\vec q})
  =& \int_{-\infty}^\infty {\rm d}x {\rm e}^{-{\rm i}q_x x}
  {\phi_0^{(n)}}^\ast (x-q_yl^2/2) \phi_0^{(m)}(x+q_yl^2/2)\cr
         =& ({2^n \over 2^m}\cdot{m! \over n!})^{1/2}
            {\rm e}^{{\vec q}^2 l^2/4}
            [(q_y-{\rm i}q_x)l/2]^{n-m}
            L_m^{n-m}({\vec q}^2 l^2/2) . \cr
}
\eqno(A)
$$

  The collective charge excitation spectrum
is determined by the equation of motion\refto{CQp}
for the charge density operator
$$
\eqalignno{
{\rm i}{\partial \over \partial t}\rho_{\beta_1 \beta_2}({\vec q})
= &[ \rho_{\beta_1 \beta_2}, H ] \cr
= &{\Delta_{SAS} \over 2}
                 [ \delta_{\beta_2, a} \rho_{\beta_1 a}({\vec q})
                 - \delta_{\beta_1, a} \rho_{a \beta_2}({\vec q})
                 - \delta_{\beta_2, s} \rho_{\beta_1 s}({\vec q})
                 + \delta_{\beta_1, s} \rho_{s \beta_2}({\vec q}) ] \cr
& +{1 \over 2L^2} \sum_{\vec k} \sum_{\alpha_1 \alpha_2 \alpha_3 \alpha_4}
     v_{\alpha_1 \alpha_2 \alpha_3 \alpha_4}^{eff}({\vec q}) \cr
& \times
       [   \delta_{\alpha_1,\beta_2}
          {\rm e}^{-{\rm i}(q_xk_y - q_yk_x)l^2/2}
          \{ \rho_{\beta_1 \alpha_2}({\vec q}+{\vec k}),
             \rho_{\alpha_3 \alpha_4}(-{\vec k}) \} \cr
  &\ \ \     -  \delta_{\alpha_2,\beta_1}
          {\rm e}^{{\rm i}(q_xk_y - q_yk_x)l^2/2}
          \{ \rho_{\alpha_1 \beta_2}({\vec q}+{\vec k}),
             \rho_{\alpha_3 \alpha_4}(-{\vec k}) \} ] . & (motion0)
           }
$$
The equations until now are applicable for general situations.
  Hereafter we specify our discussion to the
particle-hole symmetric case\refto{YMG,MR}
where the filling factor of the electron in each layer, $\nu$, is $1/2$,
and assume the ground state is the FOS state.
Then  the only non-zero expectation value in the RPA-decoupling
is $\langle c_{ks}^{(0)^\dagger} c_{ks}^{(0)}\rangle  = 1$, and
the equation of motion is linearized to give
$$
{\rm i}{\partial \over \partial t}\langle\rho_{ss}({\vec q})\rangle  = 0 ,
\eqno(motion1)
$$
$$
{\rm i}{\partial \over \partial t}\langle\rho_{aa}({\vec q})\rangle  = 0 ,
\eqno(motion2)
$$
$$
\eqalignno{
{\rm i}{\partial \over \partial t}
 \langle\rho_{as}({\vec q}) + \rho_{sa}({\vec q})\rangle
=& [-\Delta_{SAS}+{\tilde V}_{sasa}(0)-{\tilde V}_{ssss}(0) \cr
& +{\tilde V}_{ssaa}({\vec q})-{\tilde V}_{sasa}({\vec q})]
 \langle\rho_{as}({\vec q}) - \rho_{sa}({\vec q})\rangle  , & (motion3)
           }
$$
$$
\eqalignno{
{\rm i}{\partial \over \partial t}
 \langle\rho_{as}({\vec q}) - \rho_{sa}({\vec q})\rangle
= [-\Delta_{SAS}+&{\tilde V}_{sasa}(0)-{\tilde V}_{ssss}(0)
  +{\tilde V}_{ssaa}({\vec q})-{\tilde V}_{sasa}({\vec q}) \cr
& -{2 \over 2\pi l^2} v_{sasa}^{eff}({\vec q})]
 \langle\rho_{as}({\vec q}) + \rho_{sa}({\vec q})\rangle  , & (motion4)
           }
$$
where
$$
{\tilde V}_{\alpha_1 \alpha_2 \alpha_3 \alpha_4}(q_y;k_1-k_2)
      = {F}_{\alpha_1 \alpha_2 \alpha_3 \alpha_4}(q_y;k_1-k_2)
      + { G}_{\alpha_1 \alpha_2 \alpha_3 \alpha_4}(q_y;k_1-k_2) ,
\eqno(Veff)
$$
with
$$
{ F}_{\alpha_1 \alpha_2 \alpha_3 \alpha_4}({\vec q})
= {1 \over L^2} \sum_{\vec k}
   f_{\alpha_1 \alpha_2 \alpha_3 \alpha_4}^{(0,0)}({\vec k})
   {\rm e}^{-{\rm i}(q_xk_y - q_yk_x)l^2} ,
\eqno(Feff)
$$
and
$$
{ G}_{\alpha_1 \alpha_2 \alpha_3 \alpha_4}({\vec q})
= {1 \over L^2} \sum_{\vec k}
   g_{\alpha_1 \alpha_2 \alpha_3 \alpha_4}({\vec k})
   {\rm e}^{-{\rm i}(q_xk_y - q_yk_x)l^2} .
\eqno(Geff)
$$
  With these equations
the energy of the collective charge excitation, $\omega({\vec q})$,
is obtained as
$$
\omega^2({\vec q})
=E^2({\vec q})-2E({\vec q}) \cdot V^{res}({\vec q}),
\eqno(dispersion)
$$
with
$$
E({\vec q})= \Delta_{SAS}-{\tilde V}_{sasa}(0)+{\tilde V}_{ssss}(0)
           -{\tilde V}_{sasa}({\vec q})+{\tilde V}_{ssss}({\vec q}),
\eqno(E)
$$
and
$$
V^{res}({\vec q})
={\tilde V}_{sasa}({\vec q}) - {1 \over 2\pi l^2} v_{sasa}^{eff}({\vec q}).
\eqno(Vres)
$$
  \Eq{dispersion} is similar to
that for a hard-core boson system\refto{RHV,PRU}
where $E({\vec q})$ and $V^{res}({\vec q})$ correspond to
the free dispersion of the particle
and the residual interaction among particles, respectively.
  We name $E({\vec q})$ and $V^{res}({\vec q})$
as the exciton energy and the residual interaction, respectively.
\bigskip
\noindent
{\bf 3. Results and Discussions}
\taghead{3.}
\medskip
  We have performed a numerical calculation
whose details are given in Appendix A
and here we report and discuss some results.
\par
  First we summarize the results
in the case where only contributions of the lowest Landau level
are taken into account.
  The collective charge excitation spectrum
is shown for vanishing $\Delta_{SAS}$ and finite $\Delta_{SAS}$
in \Fig{dis0} and \Fig{dis1}, respectively.
  Near the origin the spectrum has a linear dispersion
which corresponds to the phonon mode of the hard-core boson system.
  This gapless Goldstone mode greatly contrasts with
the gapful incompressible mode
of the quantum Hall effect (QHE) state:
While the low energy density fluctuation is impossible in a single layer,
it becomes possible in a double layer system as a mode
where the density of each layer fluctuates out-of-phase
under the condition that the total density is kept constant.
  For the vanishing layer separation
the residual interaction between excitons, $V^{res}(0)$, vanishes
and the dispersion becomes quadratic.
  The velocity in the long wave length limit is shown in \Fig{velocity}.
(See Appendix B.)
  In the case of $\omega_c = \infty$
where only the contribution of the lowest Landau level is taken into account
the velocity is given by\refto{F}
$$
v \simeq {\pi^{1/2} \over 4}{e^2 \over \epsilon}{d \over l},
\eqno(sound)
$$
for small layer separation
and this result is related to the ground state energy
per electron-hole pair,\refto{YM} $E_e$,
via the compressibility as ${\rm d}E_e/{\rm d}\nu \propto v^2$;
$$
{{\rm d}E_e \over {\rm d}\nu} \simeq
{e^2 \over 4\pi \epsilon l}{\pi^{1/2} \over 2^{3/2}}{d^2 \over l^2} .
\eqno(comp)
$$
  In the case of finite $\Delta_{SAS}$ gap
opens at long wave-length limit.

For any value of the \dsas
the roton structure evolves
and the softening of the excitonic roton minimum of the dispersion
is observed as the layer separation increases.
  The evolution of the roton structure
results from the increase of the residual interaction
and the softening corresponds to the decrease of the overlap
of the wave functions
between the true ground state and the excitonic state at $d=0$.\refto{YM,YMG}
  Thus the collapse of the roton minimum signals the transition
between the excitonic bose condensate
and some other ground states.
The phase boundary is shown in \Fig{boundary}
which is equivalent to that obtained
in the single mode approximation.\refto{MPB}
An excitonic Wigner crystal\refto{CQl,CQp} has been
proposed for the ground state
in the high $d$ side of the boundary.\refto{CQl,CQp}
\par
  The corrections due to contributions of higher Landau levels
are also shown in previous figures:
We take $\omega_c/(e^2/\epsilon l)$ as 5 and 2
where the former is chosen
so that our perturbative calculation should be valid
and the latter is chosen by considering the fact
that the typical value of this parameter is about 2
when $\omega_c$ is about 300K and $e^2/\epsilon l$ is about 160K
at B=10T using $\epsilon = 13$.
  The energy of the excitation is always smaller as $\omega_c$ decreases
for smaller $d$ as shown in \Fig{dis0} and \Fig{dis1}(a),(b)
but the roton minimum is pushed up near the transition
as shown in \Fig{dis1}(c).
  Then the softening of the spectrum becomes harder and
the phase boundary is reached slower for smaller $\omega_c$
as the layer separation is increased as shown in \Fig{boundary}.
  These consequences are easily understood by \Eq{dispersion} as follows.
  The residual interaction between excitons, $V^{res}({\vec q})$, is small
for small $d$, since it originates from $v_{sasa}(q)$, \Eq{vsasa}.
  Then $\omega({\vec q})$ is mainly determined
by the free dispersion, $E({\vec q})$.
  Now $E({\vec q})-\Delta_{SAS}$ is proportional to
$v^{eff}_{\alpha_1 \alpha_2 \alpha_3 \alpha_4}({\vec q})$, \Eq{veff},
which decreases as $\omega_c$ becomes smaller,
since the higher-Landau level correction,
$g_{\alpha_1 \alpha_2 \alpha_3 \alpha_4}({\vec q})$,
is negative and inversely proportional to $\omega_c$.
  Therefore the excitation energy decreases
together with $\omega_c$ for small $d$.
  On the other hand, for larger $d$,
$V^{res}({\vec q})$ begins to play a role in $\omega({\vec q})$,
which leads to the evolution of the roton structure.
  This $V^{res}({\vec q})$ also decreases with $\omega_c$,
thus the roton minimum gets shallower for smaller $\omega_c$.
\par
  In this paper we investigated the effect of the higher Landau levels on the
excitation spectrum.
We have found that the quantum Hall state becomes stabler when
$\omega_c/((e^2/\epsilon l)$ becomes smaller.
However, the shift of the boundary is not so large.
In order to obtain a quantitative result to explain the experiments\refto{BJPW}
we have to use more realistic model
where the distribution of electrons along $z$ direction is taken into account
as pointed out by MacDonald, Platzman and Boebinger.\refto{MPB}
  This also makes the momentum dependence of the dispersion weaker,
and makes the phase boundary shift in the same direction as ours.
\vfill\eject
\noindent
{\bf Acknowledgment}
\medskip
  One of the authors (O.N.) is supported
from JSPS Fellowships for Japanese Junior Scientists.
The other author (D.Y.) is partially supported by Grant-in Aid
for General Scientific Research (04640361) and by
Grant-in Aid for Scientific Research on Priority Areas ``Computational
Physics as a New Frontier in Condensed Matter Research" (04231105)
both from the Ministry of Education, Science and Culture, Japan.
\bigskip
\noindent
{\bf Appendix A}
\taghead{A.}
\medskip
  Here we give the details of the numerical calculations.
The expression, \Eq{dispersion},
for the excitation spectrum, $\omega({\vec q})$,
is expressed by
the exciton energy, $E({\vec q})$,
and the residual interaction, $V^{res}({\vec q})$.
Each of these consists of two terms:
$$
E(q)=E_0(q)+E_1(q),
\eqno(AE01)
$$
and
$$
V^{res}({\vec q})=V_0^{res}(q)+V_1^{res}(q) .
\eqno(AV01)
$$
Here $E_0(q)$ and $V_0^{res}(q)$ are contribution from the lowest
Landau levels, and involve only $f$ and ${  F}$.
On the other hand $E_1(q)$ and $V_1^{res}(q)$ give correction from
the higher Landau levels, and involve only $g$ and ${  G}$.
\par
  First we present the contribution from the lowest Landau level:
$$
E_0(q)
= \int_0^\infty {\rm d}k {e^2 \over \epsilon}{\rm e}^{-kd-k^2l^2/2}
                             [1 - J_0(qkl^2)]
\eqno(AE0)
$$
and
$$
V^{res}({\vec q})\equiv V_0^{res}({\vec q})
                 = {  F}_{sasa}({\vec q})
                 - {1 \over 2\pi l^2} f_{sasa}({\vec q})
\eqno(AV0)
$$
with
$$
{  F}_{sasa}({\vec q})
= \int_0^\infty {\rm d}k {e^2 \over 2\epsilon}[1 - {\rm e}^{-kd}]
                          {\rm e}^{-k^2l^2/2} J_0(qkl^2) .
\eqno(AF)
$$
and
$$
f_{sasa}({\vec q})
={\pi e^2 \over \epsilon q}[1 - {\rm e}^{-qd}]{\rm e}^{-q^2l^2/2}
\eqno(Af)
$$
\par
  Next we present the contribution from higher Landau levels.
Here we have to know the projected interactions,
${  G}_{\alpha_1 \alpha_2 \alpha_3 \alpha_4}({\vec q})$, \Eq{Geff} and
$g_{\alpha_1 \alpha_2 \alpha_3 \alpha_4}({\vec q})$, \Eq{g}.
Noticing that $L_0^\alpha(x)=1$ we obtain
$$
A^{(0,n)}({\vec q}) {A^{(0,n)}}^\ast({\vec q})
= {1 \over n!}{\rm e}^{{\vec q}^2 l^2/2}
  ({\vec q}^2 l^2/2)^{n} .
\eqno(AAA00)
$$
  Then the projected interaction is expressed as
$$
\eqalignno{
{  G}_{\alpha \beta \gamma \delta}(q)
=-& {1 \over \omega_c}\int_0^\infty{P{\rm d}P \over \pi}
                      \int_0^\infty{Q{\rm d}Q \over \pi}
                      {\rm e}^{-P^2l^2/2}
                      {\rm e}^{-Q^2l^2/2} \cr
&\times               v_{\alpha \beta \gamma \delta}(P)
                      v_{\alpha \beta \gamma \delta}(Q)
                      I(P,Q)
                      S(P,Q;q), &(AG)
            }
$$
with
$$
I(P,Q)=\int_0^X{{\rm d}Y \over Y}[I_0(2{\sqrt Y})-1];\ \
X=(P^2l^2/2)\cdot(Q^2l^2/2),
\eqno(AI)
$$
and
$$
S(P,Q;k)=\sum_{l=-\infty}^\infty
          J_{-l}(kPl^2)
          J_{-l}(kQl^2)
          J_{l}(PQl^2),
\eqno(AS)
$$
where we have used the formula
$$
\sum_{n=1}^\infty{1 \over n}{1 \over (n!)^2} X^n
= \int_0^X{{\rm d}Y \over Y}[I_0(2{\sqrt Y})-1] .
\eqno(Asum)
$$
  Thus the correction to the exciton energy is given by
$$
\eqalignno{
E_1(q)/(e^2/\epsilon l)
={2 \over \omega_c}& \int_0^\infty {\rm d}Pl \int_0^\infty {\rm d}Ql
({\rm e}^{-Pd}+{\rm e}^{-Qd}){\rm e}^{-P^2l^2/2}{\rm e}^{-Q^2l^2/2} \cr
&\times I(P,Q)[S(P,Q;q) - J_0(PQl^2)] .  &(AE1)
           }
$$
The correction to the residual interaction between excitons is given by
$$
V_1^{res}(q)/(e^2/\epsilon l)
=G_1(q)+g_1(q),
\eqno(AV1)
$$
where
$$
\eqalignno{
G_1(q)=-{1 \over \omega_c}\int_0^\infty {\rm d}Pl \int_0^\infty {\rm d}Ql
       &(1-{\rm e}^{-Pd})(1-{\rm e}^{-Qd})
       {\rm e}^{-P^2l^2/2}{\rm e}^{-Q^2l^2/2} \cr
       &\times I(P,Q) S(P,Q;q), &(AG1)
           }
$$
and
$$
g_1(q)={1 \over \omega_c}{\rm e}^{-q^2l^2/4}
       \int_0^\infty {p{\rm d}p \over 2\pi}
       {\rm e}^{-p^2l^2/4}T(p,q).
\eqno(Ag1)
$$
Here
$$
\eqalignno{
T(p,q)=&\int_0^{2\pi} {{\rm d}\theta \over 2\pi}
        {\rm cos}[{pql^2 \over 2}{\rm sin}\theta]
v_{sasa}({1 \over 2}\sqrt{p^2+q^2+2pq{\rm cos}\theta}) &(AT) \cr
&\times   v_{sasa}({1 \over 2}\sqrt{p^2+q^2-2pq{\rm cos}\theta})
        I(p,q;\theta),
            }
$$
with
$$
\eqalignno{
I(p,q;\theta)=&\int_0^X{{\rm d}Y \over Y}[I_0(2{\sqrt Y})-1]; &(AII)
\cr
            X=&(p^2+q^2+2pq{\rm cos}\theta)l^2/8 \cdot
               (p^2+q^2-2pq{\rm cos}\theta)l^2/8 .
            }
$$
The integrals here are done numerically, and we obtain the spectrum.
\bigskip
\noindent
{\bf Appendix B}
\taghead{B.}
\medskip
  In the long wave length limit
the dispersion relation, \Eq{dispersion}, reduces to
$$
\omega^2(q) \simeq
-2[E_0^{(2)}+E_1^{(2)}]V^{res}(0)q^2l^2
\equiv v^2 q^2,
\eqno(Bdispersion)
$$
where $E_0^{(2)}$, the contribution from the lowest Landau level,
 and $E_1^{(2)}$, the contribution from the higher levels, are given by
$$
E_0(q)/(e^2/\epsilon l)
\simeq E_0^{(2)} q^2l^2
= {q^2l^2 \over 4} \int_0^\infty {\rm d}kl k^2l^2
       {\rm e}^{-kd-k^2l^2/2},
\eqno(BE0)
$$
and
$$
\eqalignno{
E_1(q)/(e^2/\epsilon l)
&\simeq E_1^{(2)} q^2l^2 \cr
&= -{q^2l^2 \over 2\omega_c}
\int_0^\infty{\rm d}Pl \int_0^\infty{\rm d}Ql
        ({\rm e}^{-Pd}+{\rm e}^{-Qd}){\rm e}^{-P^2l^2/2}{\rm e}^{-Q^2l^2/2}
\cr
   &~~~\times I(P,Q) J_0(PQl^2) (P^2+Q^2)l^2 ,
&(BE1)
           }
$$
with $E_0(q)$ and $E_1(q)$ being defined in \Eq{AE0} and \Eq{AE1},
respectively.
\vfill\eject
\def\PRL{Phys.\ Rev.\ Lett.\ }
\def\PR{Phys.\ Rev.\ }
\def\JPSJ{J.\ Phys.\ Soc.\ Jpn.\ }
\noindent
\references

\refis{CQl} X.\ M.\ Chen and J.\ J.\ Quinn:
            \PRL {\bf 67} (1991) 895.

\refis{CQp} X.\ M.\ Chen and J.\ J.\ Quinn:
            \PR {\bf B45} (1992) 11054.

\refis{PRU} D.\ Paquet, T.\ M.\ Rice and K.\ Ueda:
            \PR {\bf B32} (1985) 5208.

\refis{MPB} A.\ H.\ MacDonald, P.\ M.\ Platzman and G.\ S.\ Boebinger:
            \PRL {\bf 65} (1990) 775.

\refis{YM} D.\ Yoshioka and A.\ H.\ MacDonald:
           \JPSJ {\bf 59} (1990) 4211.

\refis{F} H.\ A.\ Fertig:
          \PR {\bf B40} (1989) 1087.

\refis{B} L.\ Brey:
          \PRL {\bf 65} (1990) 903.

\refis{HXDSZ} S.\ He, X.\ C.\ Xie, S.\ Das Sarma and F.\ C.\ Zhang:
              \PR {\bf B43} (1991) 9339.

\refis{YMG} D.\ Yoshioka, A.\ H.\ MacDonald and S.\ M.\ Girvin:
            \PR {\bf B39} (1989) 1932.

\refis{MR} A.\ H.\ MacDonald and E.\ H.\ Rezayi:
           \PR {\bf B42} (1990) 3224.

\refis{WZ} X.\ G.\ Wen and A.\ Zee:
           \PRL {\bf 69} (1992) 1811.


\refis{BJPW} G.\ S.\ Boebinger, H.\ W.\ Jiang,
             L.\ N.\ Pfeiffer and K.\ W.\ West:
             \PRL {\bf 64} (1990) 1793.

\refis{SJSSG} M.\ Shayegan, J.\ Jo, Y.\ W.\ Suen,
              M.\ B.\ Santos and V.\ J.\ Goldman:
              \PRL {\bf 65} (1990) 2916.

\refis{SJSEHS} Y.\ W.\ Suen, J.\ Jo, M.\ B.\ Santos,
               L.\ W.\ Engel, S.\ W.\ Hwang and M.\ Shayegan:
               \PR {\bf B44} (1991) 5947.

\refis{SESST} Y.\ W.\ Suen, L.\ W.\ Engel, M.\ B.\ Santos,
              M.\ Shayegan and D.\ C.\ Tsui:
              \PRL {\bf 68} (1992) 1379.

\refis{EBPWH} J.\ P.\ Eisenstein, G.\ S.\ Boebinger, L.\ N.\ Pfeiffer
              K.\ W.\ West and S.\ He:
              \PRL {\bf 68} (1992) 1383.

\refis{SJSES} Y.\ W.\ Suen, J.\ Jo, M.\ B.\ Santos,
              L.\ W.\ Engel and M.\ Shayegan:
              to be published in Physica {\bf B}
              (Proceedings of the Yamada Conference on the Application
              of High Magnetic Fields in Semiconductor Physics,
              Chiba, Japan, August 1992).

\refis{KMH} R.\ Kubo, S.\ J.\ Miyake and N.\ Hashitsume:
            in {\it Solid State Physics,} ed. F. Seitz and D. Turnbull
            (Academic, New York, 1965), Vol. 17, p269.

\refis{RPM} M.\ Rasolt, F.\ Perrot and A.\ H.\ MacDonald:
            \PRL {\bf 55} (1985) 433.

\refis{RHV} M.\ Rasolt, B.\ I.\ Halperin and D.\ Vanderbilt:
            \PRL {\bf 57} (1986) 126.

\endreferences
\vfill\eject
\noindent
\figurecaptions

\figis{dis0}
       Dispersion of collective charge excitation
       for $\Delta_{SAS}=0$ with $d/l=0.5$ (a) and $d/l=1.1$ (b).
       The solid line represents the result
       only with contributions of the lowest Landau level
       ($\omega_c/(e^2/\epsilon l)=\infty$).
       The broken and dotted lines represent the results
       with contributions of higher Landau levels
       for $\omega_c/(e^2/\epsilon l)=5$ and 2, respectively.

\figis{dis1}
       Dispersion of collective charge excitation
       for $\Delta_{SAS}/(e^2/\epsilon l)=0.2$
       with $d/l=0.5$ (a), $d/l=1.1$ (b) and $d/l=2.3$ (c).

\figis{boundary}
       Phase boundary of the quantum Hall state
       signaled by the softening of the roton minimum of the dispersion
       as a function of $d/l$ and $\Delta_{SAS}/(e^2/\epsilon l)$ .
       The quantum Hall state is stable in the lower side of the boundary and
       is unstable in the upper side.

\figis{velocity}
       Velocity in the long wave length limit
       for $\Delta_{SAS}=0$ as a function of $d/l$ .

\endfigurecaptions

\end